\documentclass[apj]{emulateapj}
\usepackage{apjfonts}
\usepackage{amsmath}
\newcommand{\Kp}{$K^{\prime}$}

\slugcomment{Accepted for Publication in ApJ}
\newcommand{\lya}{\mbox{${\rm Ly}\alpha$}}
\newcommand{\etal}{\ensuremath{\mbox{et~al.}}}
\providecommand{\kms}{\,\ensuremath{\rm{km\,s}^{-1}}}

\newcommand{\apll}{\lesssim}
\newcommand{\asec}{$''$}
\newcommand{\amin}{$'$}

\shorttitle{Mg\,II Galaxies Revealed by GRB\,060418}

\shortauthors{Pollack \etal}

\begin{document}


\title{An Imaging and Spectroscopic Study of Four Strong Mg\,II Absorbers Revealed By GRB\,060418}

\author{L. K. Pollack\altaffilmark{2}, H.-W. Chen\altaffilmark{3},
J. X. Prochaska\altaffilmark{2,4}, \& J. S. Bloom\altaffilmark{5,6}}

\altaffiltext{1}{Based in part on observations made with the NASA/ESA
Hubble Space Telescope, obtained at the Space Telescope Science
Institute, which is operated by the Association of Universities for
Research in Astronomy, Inc., under NASA contract NAS 5-26555. }

\altaffiltext{2}{Department of Astronomy \& Astrophysics, University
of California, Santa Cruz 95064} 

\altaffiltext{3}{Dept.\ of Astronomy \& Astrophysics and Kavli
Institute for Cosmological Physics, University of Chicago, Chicago,
IL, 60637, U.S.A. \\ {\tt hchen@oddjob.uchicago.edu}}

\altaffiltext{4}{University of California Lick Observatories}

\altaffiltext{5}{Department of Astronomy,
        University of California, Berkeley, CA 94720-3411.}
\altaffiltext{6}{Sloan Research Fellow.}

\begin{abstract}

We present results from an imaging and spectroscopic study of four
strong Mg\,II absorbers of $W(2796) \gtrsim 1$ \AA\ revealed by the
afterglow of GRB\,060418 at $z_{\rm GRB}=1.491$.  These absorbers, at
$z=0.603,0.656,1.107$ and $z_{\rm GRB}$, exhibit large ion abundances
that suggest neutral gas columns characteristic of damped \lya\
systems.  The imaging data include optical images obtained using LRIS
on the Keck I telescope and using ACS on board HST, and near-infrared
$H$-band images obtained using PANIC on the Magellan Baade Telescope
and $K'$-band images obtained using NIRC2 with LGSAO on the Keck II
telescope.  These images reveal six distinct objects at
$\Delta\,\theta\apll 3.5''$ of the afterglow's position, two of which
exhibit well-resolved mature disk morphology, one shows red colors,
and three are blue compact sources.  Follow-up spectroscopic
observations using LRIS confirm that one of the disk galaxies
coincides with the Mg\,II absorber at $z=0.656$.  The observed
broad-band spectral energy distributions of the second disk galaxy and
the red source indicate that they are associated with the absorbers at
$z=0.603$ and $z=1.107$, respectively.  These results show that strong
Mg\,II absorbers identified in GRB afterglow spectra are associated
with typical galaxies of luminosity $\approx 0.1-1\,L_*$ at impact
parameter of $\rho \apll 10\ h^{-1}$ kpc.  The close angular
separation would preclude easy detections toward a bright quasar.
Finally, we associate the remaining three blue compact sources with
the GRB host galaxy, noting that they are likely star-forming knots
located at projected distances of $\rho=2-12\ h^{-1}$ kpc from the
afterglow.  At the afterglow's position, we derive a 2-$\sigma$ upper
limit to the underlying SFR intensity of 0.0074 M$_\odot$ yr$^{-1}$
kpc$^{-2}$.

\end{abstract}

\section{Introduction}

Quasars (QSOs) have long been exploited as bright, continuum sources
with absorption-line spectra that can reveal the presence of
intervening gas at high redshifts \citep{rauch97,wgp05}.
A subset of these absorbing systems offers the opportunity to study
the halo gas and interstellar medium (ISM) of galaxies at a large
range of evolutionary stages, independent of stellar luminosity.
Spectroscopy of the optical afterglows of gamma-ray bursts (GRBs) has
recently become an exciting alternative to quasar absorption line
studies.  The transient nature of GRB afterglows enables a more
sensitive and systematic search for the stellar component of the GRB
host, as well as any absorbing systems serendipitously aligned along
that sightline.  In contrast with quasar studies (where photons from
the central source have ample time to illuminate the gas and dust
along the line of sight), this presents an opportunity to compare the
stellar and gaseous properties of high-redshift galaxies at a variey
of galactocentric radii.

From detailed studies of gaseous properties alone, much insight has
been gained about galaxy formation and gas dynamics in the early
universe.  \citet{Prochaska:1997} challenged the standard picture of
galaxy formation \citep[e.g.][]{Kauffmann:1996, Mo:1998} when they
demonstrated that the neutral gas in damped Ly$\alpha$ (DLA) systems
of $\log\,N({\rm H\,I})\ge 20.3$ \citep[e.g.][]{wgp05} exhibited
kinematics favoring rotational dynamics, but with velocity widths too
large to have arisen from overdensities in a universe filled with cold
dark matter.  Subsequent numerical simulations revealed that velocity
fields from infall, accretion, and turbulence may also contribute to
apparent rotation in the merging clumps of protogalaxies
\citep{Haehnelt:1998}.  In addition to motions related to
gravitational dynamics, some models of quasar absorption line systems
(e.g.\ \ion{Mg}{2} systems) invoke superwinds and outflows, especially
to describe the most extreme velocity fields observed
\citep{nbf98,Bond:2001,Bouche:2007}.

While contributing significantly to our understanding of gas dynamics
in and about galaxies, these past works share one commonality: they
draw conclusions almost entirely from absorption properties.  The
primary difficulty is in associating a certain absorption spectrum
with the appropriate galaxy, as seen in emission.  Indeed, without the
knowledge of a system's stellar properties such as the systemic
velocity, which is usually measured from nebular emission lines, it is
difficult to distinguish between models involving the organized motion
of outflows \citep{Dong:2003,Cox:2006} and models that include
virialized motions and gravitational accretion
\citep{Mo:1996,McDonald:1999}.  Nevertheless, without a robust
connection of a particular galaxy to a particular absorption spectrum,
one cannot test claims of starbursts and superwinds against a galaxy's
morphology, luminosity, and color.

Such studies, however, are just beginning to be explored.  While {\bf
$\sim 1000$} DLAs are known to exist, only a handful have been found
\citep{Moller:2002,OMeara:2006} at redshifts $z>1$.  At $z<1$, a
larger sample of known DLA absorbing galaxies has been established
\citep{rnt+03,Chen:2003}, although roughly 40\% of the low-redshift
DLA population remain unidentified with their stellar counterpart.
Likewise, the nature of the galaxies associated with strong Mg\,II
absorption systems ($W(2796)>1$ \AA) detected along quasar sightlines
has recently been scrutinized with a similar success rate
\citep[e.g.][]{Bouche:2007}. Historically, the presence of a bright
quasar has made it exceedingly difficult to detect the stellar
counterparts of high-column-density absorbers along the sightline.
The sample of known absorbing galaxies is inherently biased toward
higher luminosities and higher impact parameters with a median angular
distance to the background QSO of $\langle\theta\rangle=3.6''$ for
known DLA galaxies \cite[[e.g.][]{Chen:2003} and
$\langle\theta\rangle=2.4''$ for strong Mg\,II absorbing galaxies
\cite[[e.g.][]{Bouche:2007}.  This inhearent bias makes the absorbers
notoriously difficult to unambiguously identify.  Yet these systems
are crucial to our understanding of the baryonic content of galactic
halos, and can help discriminate between competing scenarios for the
nature of extended gas \citep[e.g.][]{CT:2008}.  The use of a
transient GRB afterglow as the bright background source is one remedy
to these past difficulties.  The {\it Swift} satellite
\citep{Gehrels:2004} has thus revolutionized the study of
high-redshift starforming galaxies, having detected approximately 300
GRBs, of which {\bf $\sim 120$} were localized through their optical
transients.

Immediate spectroscopic and imaging follow-up campaigns, as well as
late-time deep imaging observations, are equally crucial to fully
leverage such a fortuitous occurence.  In recent years, dedicated
campaigns have resulted in significant advancements in our
understanding of the progenitors and host galaxies of GRBs.
Early-time, spectroscopic follow-up observations have revealed the
detailed chemical abundances and approximate dust contents and star
formation histories of GRB hosts, through measurements of abundance
ratios such as $[\alpha/\rm Fe]$ and $[\rm Ti/Fe]$
\citep{Savaglio:2006,Prochaska:2007}.  In addition, analysis of the
time evolution of Fe\,II and Ni\,II excited and metastable populations
has proven to be an excellent measure of the distance to the neutral
gas along the sightline \citep{pcb06,dcp+06,Vreeswijk:2007}; such
analysis makes use of the UV-pumping of the $\sim$\,kpc environment by
the GRB itself to unambiguously identify the redshift of the host
galaxy.  Studies of the morphologies of GRB host galaxies, and of the
offsets between GRB afterglows and the light of the putative host
galaxy \citep{Bloom:2002} have hinted that GRB progenitors may be more
massive than the progenitors of core-collapse supernova
\citep{Fruchter:2006,Kelly:2008}.

However, we point out that much of the research done on GRB hosts to
date has required the automation and high spatial-resolution afforded
by the HST \citep{Bloom:2002,cgh+03,Jakobsson:2003,fgs+05,Chen:2009}.
The aforementioned investigation, which utilized HST observations
taken at various times after the burst, required that the GRB
afterglow positions were known with excellent accuracy (less than a
few tenths arcsecond).  In contrast, the {\it Swift}/UVOT regularly
provides afterglow position estimates with uncertainties on the order
of 0.5\asec -- equivalent to the half-light radii typical of
high-redshift galaxies.  Thus UVOT position estimates, while adequate
for identifying the probable host, are inadequate for analyses of the
burst's local environment, or studies that require accurately known
galactocentric radii.  Both the shortfalls of {\it Swift} and the high
demand for HST time underline the need for a new strategy of
ground-based follow-up observations that can localize the afterglow
with high accuracy.

In this work we demonstrate such a strategy.  We employ a combination
of low-resolution, early-time, automated observations from a 1\,m
telescope, and high-resolution, deep, late-time observations from
$8-10$\,m class telescopes, including laser guide star adaptive optics
imaging.  The data resulting from this combination of observations is
exceptionally rich, and allows us to study the GRB's host galaxy as
well as any foreground absorbers along the afterglow's sightline.  We
can study the morphologies and stellar properties of the host and
foreground galaxies at near-infrared wavelengths with spatial
resolutions comparable to the HST, and we can interpret these
properties while knowing the galaxy's gas contents at accurate
galactocentric radii.

To showcase this strategy, which fuses the early-time results from
teams like GRAASP\footnote{Gamma-Ray Burst Afterglows As Probes;
http://www.graasp.org/} with long-term
follow-up campaigns, we have analyzed the rich sightline toward
GRB\,060418 at $z=1.491$.  This sightline has the potential to
contribute significantly to our understanding of Mg\,II absorption
systems, as three intervening ($z=0.6-1.1$), strong Mg\,II absorbers
of rest-frame absorption equivalent width $W(2796) \ge 1$ \AA\ were
discovered in the early-time high-resolution spectroscopic
observations of the afterglow \citep{Ellison:2006, Prochaska:2007b}.
While there is a known overabundance of Mg\,II systems along GRB
sightlines compared to those along quasar sightlines
\citep{Prochter:2006}, a triplet of foreground Mg\,II absorbers is
rarely observed along a single sightline and offers a
unique opportunity to study the extended gas of three
galaxy halos.  In addition, the observed overabundance of \ion{Mg}{2}
absorbers remains
unexplained, and partially motivated this work.  Despite this
overabundance, and despite the relative ease of identifying Mg\,II
systems along GRB sightlines, only two spectroscopic confirmations of
Mg\,II absorbers have ever been made toward GRBs
\citep{Jakobsson:2004,Masetti:2003}.  In this paper we announce the
third such spectroscopic confirmation.

Finally, although we lack an HI column density measurement for the
host galaxy of the burst and for the foreground absorbers\footnote{The
Ly$\alpha$ line is blueward of the atmospheric cutoff at redshifts
below $z=1.6$.}, the large ion abundances observed in the Mg\,II
absorbers suggest that these are likely DLAs.  For example, the Mg\,II
absorber at $z=0.603$ contains $\log\,N({\rm Fe\,II})=15.67$
\citep{Prochaska:2007b}.  A solar abundance of the absorbing gas
without dust depletion would imply an underlying total neutral gas
column density of $\log\,N({\rm H\,I})=20.2$, and a 0.1 solar
metallicity would imply $\log\,N({\rm H\,I})=21.2$.  While we expect
the GRB host to be a DLA galaxy, the foreground DLAs along this GRB
sightline are fortuitous, and can be compared directly to the
traditional DLA population discovered along quasar sightlines (termed
QSO-DLAs).  Thus, this one sightline adds appreciably to the small
number of high-redshift DLAs with known stellar counterparts.

Throughout this paper we use $\Omega_{M}=0.3$ and
$\Omega_{\lambda}=0.7$, and unless otherwise stated we adopt
$H_{0}=75\,\rm km\,s^{-1}\,Mpc^{-1}$.

\section{Observations}

\subsection{Imaging Observations}

\begin{deluxetable*}{lcccc}
\tablecaption{\sc{Journal of Imaging Observations} \label{tab:Log}}
\tablewidth{0pt}
\tablehead{\colhead{Telescope/Instrument} & \colhead{Filter} & \colhead{Total Exposure Time} & \colhead{Mean FWHM} & \colhead{UT Date}}
\startdata
Keck/LRIS & $g$ & 30 min        & 1.1$''$ & 30 May 2006 \\
Keck/LRIS & $R$ & 10 min        & 1.0$''$ & 30 May 2006 \\
Magellan/PANIC & $H$ & 78 min   & 0.3$''$ & 19 May 2006 \\
PAIRITEL & J & 300 sec        & 2.8$''$ & 18 April 2006 \\
Keck/NIRC2 LGSAO & $K'$ & 23 min & 0.08$''$ & 22 June 2007 \\
HST/ACS & F555W & 4386 sec    & 0.1$''$ & 12 July 2006 \\
HST/ACS & F625W & 8772 sec    & 0.1$''$ & 11 July 2006 \\
HST/ACS & F775W & 8772 sec    & 0.1$''$ & 12 July 2006 \\
\enddata
\end{deluxetable*}

\subsubsection{Early-Time Swift and PAIRITEL}

On 18 April 2006 at 3:06:08 UT the {\it Swift} Burst Alert Telescope
(BAT) triggered on GRB 060418 \citep{Falcone:2006}, providing a
3\amin\, error circle localization centered at (RA, DEC) of
$\alpha$(J2000)$ = $15\,h 45\,m 41\,s, $\delta$(J2000)$ = -$03\,d 38\amin\ 35\asec.
About one minute later observations with the {\it Swift} X-Ray
Telescope (XRT) resulted in an initial position estimate with a
5.8\asec\ error centered at $\alpha$(J2000)$ = $15\,h 45\,m 42.8\,s, $\delta$(J2000)$ =
-$03\,d 38\amin\ 26.1\asec.  This position was later refined by the
{\it Swift} XRT team, resulting in a 4\asec\ error centered at $\alpha$(J2000)$
=$15\,h 45\,m 42.4\,s, $\delta$(J2000)$ = -$03\,d 38\amin\ 22.8\asec
\citep{Falcone:2006a}.  Both XRT positions are consistent with the
original BAT localization.  An afterglow with $V \sim 14.5$\,mag was
quickly discovered by the {\it Swift} UltraViolet and Optical
Telescope (UVOT).  The reported afterglow position is at the edge of
the revised XRT error circle, at $\alpha$(J2000)$ = $15\,h 45\,m 42.60\,s,
$\delta$(J2000)$ = -$03\,d 38\amin\ 20.0\asec with a $1\sigma$ error radius of
about 0.5\asec\ \citep{Falcone:2006}.

These {\it Swift} detections triggered the PAIRITEL 1.3\,m automated
telescope \citep{Bloom:2006} to slew toward GRB\,060418 at 05:25:34
UT, where it began observations of the field, obtaining 13 300\,s JHKs
mosaics, acquired simultaneously in all three bands.  The transient
was well detected in the 300\,s mosaics and faded over the next 6
hours of observations \citep{Kocevski:2006}.  We show one of the
earlier 300\,s $J$-band mosaics in Fig.~\ref{fig:groundPanel}, zoomed
in on the afterglow position. The mean seeing was 2.8\asec\ FWHM at
$J$ band, and thus the dithered images were Nyquist sampled by the
1\asec\ pixels in the reconstructed mosaics.

\subsubsection{Magellan $H$-band Images Using PANIC}

Follow-up observations of the field around GRB\,060418 were performed
on 19 May 2006 with the Persson's Auxiliary Nasmyth Infrared Camera
\citep[PANIC;][]{Martini:2004} on the Magellan Baade telescope.  We
acquired twenty-six sets of three 60\,s $H$-band images for a total
exposure time of 78 minutes.  The images were processed using standard
techniques.  Each set of three 60\,s images were initially median
combined; the resulting 26 frames were registered to a common origin,
filtered for deviant pixels, and stacked to form a final combined
image using our own software.

Although the night was not photometric, the seeing quality was
excellent, yielding a mean FWHM of 0.3\asec.  Photometric calibrations
were performed using additional images of the same field obtained on
27 August 2007 under photometric conditions.  The later data were
calibrated to Persson standards \citep[9172 and
9164;][]{Persson:1998}.  We show a portion of the final, stacked
$H$-band image in Fig.~\ref{fig:groundPanel}.  At least three objects,
all within $3$\asec\ of the afterglow position, are visible.  We name
these three objects $A$, $B$, and $C$ from north to south, and refer
to them throughout the text.  In \S~\ref{sec:absorbers} we describe
the photometric and spectroscopic properties of galaxies $A$, $B$, and
$C$, and discuss their association with known Mg\,II absorbers
previously identified in early-time afterglow spectra.

\subsubsection{Keck $g$ and $R$ Images Using LRIS}

On 30 May 2006 we obtained additional imaging observations of the
field surrounding GRB\,060418 using the Low Resolution Imaging
Spectrometer \citep[LRIS;][]{Oke:1995} on Keck I.  We took three
600\,s exposures at $g$ band for a total integration time of 30
minutes, and we took one 10 minute exposure at $R$ band.  The seeing
conditions yielded FWHMs of 1.1\asec\ and 1.0\asec\ at $g$ and
$R$ bands, respectively.  These images were processed using standard
techniques. The data were bias subtracted, flat fielded with dome
flats, and calibrated with a Landolt standard field
\citep[PG2213;][]{Landolt:1992}.  In Fig.~\ref{fig:groundPanel} we
present a 13\asec\ square region of the $R$-band image showing the
region of interest.

\subsubsection{Keck \Kp\ Using NIRC2 LGSAO}

\begin{figure*}
\begin{center}
\includegraphics[scale=.35]{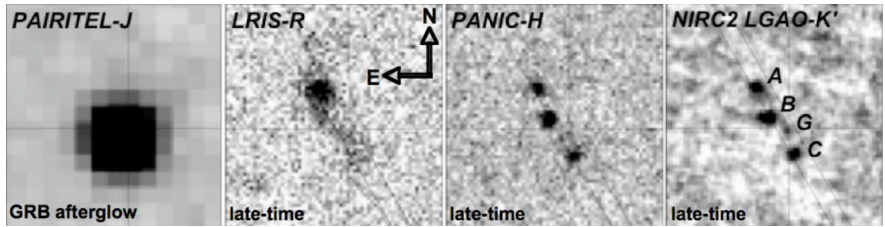}
\caption{Astrometrically aligned images, 13\asec\ on a side of the
  field surrounding GRB~060418.  North is up and East is to the left.
  From left to right, the images were taken with the following
  instruments and filters: PAIRITEL $J$, LRIS $R$, PANIC $H$, and
  NIRC2 \Kp.  The three right panels are late-time observations of the
  field, i.e.\ after the afterglow has faded to fainter than $R=26$.
  The leftmost panel shows the afterglow at $\approx$3\,hr after the
  GRB trigger.  Cross-hairs guide the eye and mark the position of
  object $G$ in the NIRC2 image.  This image has been smoothed to
  accentuate the detection of object $G$, which we believe to be part of
  the GRB's host galaxy complex.  Galaxies $A$, $B$, and $C$ are discussed
  throughout the text, and are believed to be the three intervening
  Mg\,II absorbers detected in early-time spectroscopic observations of
  the afterglow.  The orientation of the follow-up LRIS 1\asec\
  longslit (PA=30 deg) observations is also indicated.}
\label{fig:groundPanel}
\end{center}
\end{figure*}

Late-time follow-up observations were performed more than a year after
the burst, on 22 June 2007, using laser guide star adaptive optics
\citep[LGSAO;][]{Wizinowich:2006} with the NIRC2 \citep{Matthews:1994} wide
field camera (0.04\asec/pixel) on Keck II.  To maximize the camera
efficiency and avoid sky saturation, we used three coadds and exposed
for 20 seconds for a total integration time of 23 minutes at \Kp\
band.  When using the wide field camera the diffraction limited PSF
core is slightly undersampled at \Kp\ band, however the larger field
of view was chosen to include a nearby 2MASS star (21\asec\
southeast).  This star was especially valuable for image alignment,
photometric calibration, and PSF estimation.  The images were
processed using standard infrared imaging techniques, including dark
subtraction, flat fielding, and filtering for bad pixels.  Before
performing image alignment, we dewarped individual reduced frames
using the prescription outlined in the NIRC2 online
documentation.\footnote{http://www2.keck.hawaii.edu/inst/n2TopLev/post\_observing/dewarp}
The mean FWHM of the AO-corrected PSF core is 0.08\asec.  A portion of
the LGSAO image is shown in Fig.~\ref{fig:groundPanel}.  Galaxies $A$,
$B$, and $C$ are clearly detected and coincident in position with what
is seen in the coadded PANIC $H$-band image.  We note a forth object,
labeled $G$, seen at the $\sim 2\,\sigma$ level between $B$ and $C$ on
the plane of the sky.  In \S~\ref{sec:host} we discuss the photometric
properties of source $G$.  We interprete this object to be part of the
GRB host galaxy.  Throughout the text we will use the letter $G$ to
indicate that an object is physically associated with the GRB host
galaxy.

\subsubsection{Optical Images in the HST/ACS Data Archive}

\begin{figure}
\begin{center}
\includegraphics[scale=0.35]{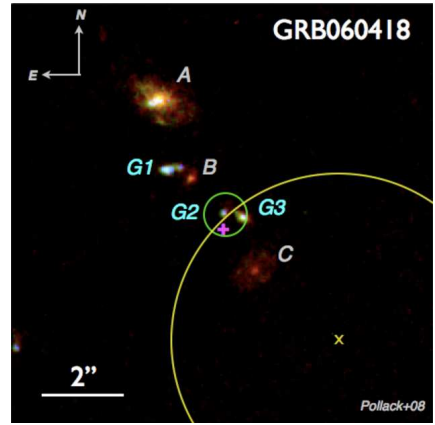}
\caption{A 10\asec\ $\times$ 10\asec\ late-time (July 2006) HST/ACS
  composite image of the field surrounding GRB~060418 made using the
  F555W, F625W, and F775W filters.  The pink plus sign indicates the
  actual position of the optical transient, as determined by alignment
  of our early-time ground based PAIRITEL data, and confirmed by the
  fading transient detected in earlier time ACS images.  The yellow
  `x' and surrounding yellow circle mark the revised XRT position and
  90\% containment radius \citep{Falcone:2006a}.  The green circle
  indicates the UVOT position and associated $1\sigma$ error radius of
  0.5\asec\ \citep{Falcone:2006}.  Objects $A$, $B$, and $C$ are
  marked for comparison with Fig.~\ref{fig:groundPanel}; these objects
  are thought to be the stellar counterparts of three intervening
  absorbers detected in the early-time afterglow spectrum.  Objects
  $G1$, $G2$, and $G3$ are all thought to be associated with the GRB
  host galaxy.}
\label{fig:ACS}
\end{center}
\end{figure}

Between May and July of 2006, multiple observations of GRB 060418 were
performed with the Hubble Space Telescope (HST) Advanced Camera for
Surveys (ACS) using the F555W, F625W, and F775W filters (PID 10551;
PI: S. Kulkarni).  The imaging data are retrieved from the HST data
archive and processed using the standard reduction pipeline.  The
first epoch of observations were carried out three weeks after the
burst.  The images show the fading afterglow which is absent in the
images taken on 11 and 12 July 2006.  The early-time observations
confirm the astrometric position of the afterglow relative to other
nearby galaxies (see \S~\ref{sec:astrometry}).  The late-time
observations complement our own ground based data in both frequency
and resolution, and provide a view of the host galaxy and intervening
systems that is unimpeded by the bright afterglow.  The mean FWHM of
the PSF for all three filters is approximately 0.1\asec.  In
Fig.~\ref{fig:ACS} we present a composite late-time image made from
all three ACS filters, registered using stars in the field.  The
position of the faded afterglow is indicated with a pink plus sign.
The three redder objects previously detected in our $H$ and \Kp\
images (objects $A$, $B$, and $C$) are again labeled on the ACS image.
In addition, we identify three blueish objects, $G1$, $G2$, and $G3$,
which we interpret to be components of the GRB host galaxy (see
\S~\ref{sec:host}).

In Table~\ref{tab:Log} we provide a journal of all imaging
observations discussed in this section.

\subsection{Spectroscopy}

We performed spectroscopic observations with the LRIS 1\asec\ longslit
on UT 16 August 2007 in the hopes of identifying the galaxies
responsible for the GRB and intervening \ion{Mg}{2} absorption along
the sightline.  Using the 680 dicrhoic we simultaneously took two
1830\,s exposures through the 300/5000 grism ($300\,\rm grooves
\,mm^{-1}$, $\lambda_{\rm blaze}=5000\,\rm \AA$), and two 1800\,s
exposures through the 600/7500 grating ($600\, \rm grooves \,mm^{-1}$,
$\lambda_{\rm blaze}=7500\,\rm \AA$), for a total integration time of
approximately 1 hour over the wavelength range $2300-9000\,\rm \AA$.
(Transmission through the 300/5000 grism drops substantially below
4000\AA.)  The data were processed with an IDL package customized for
LRIS longslit reductions developed by J. Hennawi and
J.X.P.\footnote{http://www.ucolick.org/$\sim$xavier/LowRedux}.  The
reduction includes standard bias subtraction and flat fielding,
wavelength calibration using arc lamp observations taken the same
night, and instrument flexure correction using cross-correlations of
observed and archived sky spectra.  The spectra were calibrated to
vacuum wavelengths, corrected for the heliocentric motion, and fluxed
using the spectrophotometric standard, BD33 2642.  We did not correct
the spectra for Galactic extinction, which is $E(B-V)=0.22$\,mag along
this direction \citet{Schlegel:1998}.

Using a position angle of 30 degrees (East of North), we oriented the
slit such that it would include light from all three nearby objects
that had been discovered in emission prior to the date of these
spectroscopic observations (objects $A$, $B$, and $C$).\footnote{The
proprietary period for the HST ACS observations ended in November
2007, after these spectroscopic observations were carried out.}  In
Fig.~\ref{fig:groundPanel} we show the orientation of the slit
relative to our ground-based data.  Unfortunately, the slit did not
include flux from the blueish object revealed by ACS and marked $G1$
in Fig.~\ref{fig:ACS}.  We now interpret $G1$ to be a component of the
GRB host (see \S~\ref{sec:host}).

\begin{figure*}
\begin{center}
\includegraphics[scale=0.25]{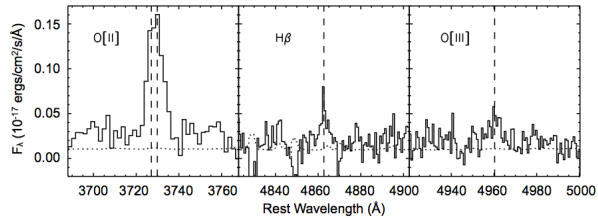}
\caption{Nebular emission lines from galaxy $A$, observed using the
  Keck/LRIS longslit.  (Galaxy $A$ is the northern-most system shown
  in Fig~\ref{fig:groundPanel}.)  The spectrum has been fluxed, but we
  did not correct for Galactic extinction.  The [\ion{O}{2}], H$\beta$, and
  [\ion{O}{3}] emission lines indicate a systemic redshift of $z=0.6554 \pm
  0.0002$.  Vertical dashed lines guide the eye showing the rest
  wavelengths of these transitions. The observed emission redshift
  corresponds to a velocity offset of 109\,km\,s$^{-1}$ blueward of
  the MgII absorption line revealed in an early-time afterglow
  spectrum \citep[$z=0.6560$,][]{Ellison:2006,Prochaska:2007b}.  The
  dotted line traces the $1\sigma$ error estimate.}
\label{fig:Longslit}
\end{center}
\end{figure*}

Continuum from just one galaxy was detected in these spectroscopic
data, located at the position of object $A$.  The spectrum of this
object is shown in Fig.~\ref{fig:Longslit}.  Nebular emission lines of
[O\,II]$\,\lambda\,3727$, H$\beta\,\lambda\,4861$, and
[O\,III]$\,\lambda\,4958$ indicate that object $A$ has a redshift of
$z=0.6554 \pm 0.0002$, which corresponds to a velocity
109\,km\,s$^{-1}$ blueward of the MgII absorption signature imposed by
an intervening system reported at $z=0.6560$ \citep{Ellison:2006,
Prochaska:2007b}.  At this redshift, the [O\,III]$\,\lambda\,5007$
falls in the forest of night sky lines at 8288 \AA\ and we were unable
to identify a robust line feature from the galaxy.  The properties of
galaxy $A$ and the identities of the other two MgII absorbers will be
discussed in detail in \S~\ref{sec:absorbers}.

\section{Analysis}
\subsection{Astrometry of Ground Based Images}\label{sec:astrometry}
An accurate interpretation of the GRB host galaxy complex and all
intervening absorbers depends crucially on the accuracy of aligning
the $G$, $R$, $J$, $H$, and \Kp\ images, and the astrometric precision
in the position of the GRB afterglow relative to other objects at
small angular distances.  The large range of wavebands and spatial
resolutions afforded by each image makes this a non-trivial task.  In
order to perform relative astrometric alignment of all of our ground
based images, we first used SExtractor \citep{Bertin:1996} to identify
objects in each field.  Objects in common in two images adjacent in
wavelength space were used as inputs to the IRAF geomap package.  We
applied the astrometric solution created by geomap to the redder image
in each pair using the geotran package.  In this way we aligned pairs
of images to each other, eventually forming an alignment between the
$g$, $R$, $J$, $H$, and \Kp\ images.  The relatively large fields of
view of the PAIRITEL, LRIS and PANIC images (about 9\amin, 6\amin\ and
3\amin, respectively) included numerous common galaxies and stars,
allowing robust alignments with $\delta_{\rm RMS}\sim 0.02$\asec.  The
much smaller field of view of the NIRC2 wide-field image (about
40\asec) was just large enough to include one alignment star and one
galaxy, other than galaxies $A$, $B$, and $C$.  However these objects
were well spaced and the resulting alignment is accurate to within
$\sim 0.07$\asec.

After aligning all of the ground-based images, we performed absolute
astrometry by correlating the positions of objects in the PAIRITEL
image to sky coordinates from the 2MASS catalog\footnote{The absolute accuracy
of the 2MASS catalog is $\sigma=250$mas \citep{2MASS}.} 
using the imwcs program in
WCSTools.\footnote{http://tdc-www.cfa.harvard.edu/software/wcstools/}
The absolute astrometric position of the afterglow in the PAIRITEL
image, as determined from this alignment procedure, fell within the
0.5\asec, $1\sigma$ error radius of the reported {\it Swift} UVOT
position \citep{Falcone:2006}.  As shown in
Fig.~\ref{fig:groundPanel}, the afterglow lies closest to object $G$
in the NIRC2 image; it is located 0.5\asec\ southeast of object $G$ on
the plane of the sky between galaxies $B$ and $C$, almost along a straight
line.  This fact was later confirmed by our analysis of the early-time
archival ACS images, that show the afterglow fading during the first three
months after the burst.

The afterglow was not coincident with the centroids of galaxies $B$ or
$C$\footnote{Recall that the spectrum of galaxy $A$ shows that it is a
foreground galaxy coincident with a Mg\,II absorber at $z=0.656$.}.
If we suppose that objects $B$ or $C$ are associated with the host,
their angular distances of 1.5\asec\ and 1.25\asec\ from the afterglow
would require them to be separated from the afterglow by physical
distances of at least $\sim 12$\,kpc and 10\,kpc, respectively.  These
distances are far beyond the half-light radii of typical high-redshift
galaxies.  Furthermore, if either object $B$ or $C$ were the host
galaxy, the afterglow would be situated much farther from its host's
flux-weighted centroid than is typically observed.  \citet{Bloom:2002}
calculated a median projected offset of 0.17\asec\ between afterglows
and the flux-weighted centroids of the putative host galaxies for 20
GRBs, 16 of these afterglows are spectroscopically identified at
$z=0.008-3.418$ with a median corresponding projected distance of
$\approx 0.9$ kpc and a maximum of 6.2 kpc (corrected for
$H_{0}=75\,\rm km\,s^{-1}\,Mpc^{-1}$).  This strongly suggests that
galaxies $B$ and $C$ are not associated with the host.  Since their
chance alignment is more probable at lower redshifts, they are likely
intervening galaxies, and strong absorbers due to their small impact
parameters.  Indeed, further evidence for this scenario is given in
Section~\ref{sec:losGalaxies}.

\subsection{Galaxy Photometry}\label{sec:galPhot}

We performed photometry on every object in the ACS images within
5\asec\ of the burst -- objects $A$, $B$, $C$, $G1$, $G2$, and $G3$ in
Fig.~\ref{fig:ACS}.  First we used SExtractor on the F775W filter to
create a segmentation map.  We used a low detection threshold
(DETECT\_THRESH$=1.5$) and deblending parameter to include as much of
the light from each galaxy as possible, while still distinguishing
between nearby objects.  This segmentation map was used to sum the
flux of each object in an algorithm developed by H.-W. C., which
performs a local sky subtraction using adjacent sky pixels.  To obtain
accurate measurements of galaxy colors, we applied this same
segmentation map to sum the flux of the objects in the two other ACS
filters. 
We converted fluxes into $AB$ magnitudes using the photometric
parameters supplied by the HST header information.  The magnitudes of
each object are recorded in Table~\ref{tab:Phot}.  They have been
corrected for a rather large Galactic extinction using
$E(B-V)=0.22$\,mag, reported in \citet{Schlegel:1998}.  Where
necessary we interpolated to find the extinction in the ACS
bandpasses; the resulting values were $\rm A_{555}=0.73$\,mag, $\rm
A_{625}=0.63$\,mag, and $\rm A_{775}=0.47$\,mag.

Photometry on objects in the LRIS $g$ and $R$ and PANIC $H$ images was
less straightforward.  None of the objects $G1$, $G2$, or $G3$ are
resolved or detected in these data, and only galaxy $A$ is well
detected in the LRIS images; however knowledge of the complexity of
this field guided our photometric measurements.  By comparing the
results from standard aperture photometry using varying aperture
radii, with isophotal photometry using varying segmentation maps
created by SExtractor, we were able to identify apertures that
overestimated flux due to inclusion of nearby contaminating objects,
and underestimated flux in specific seeing conditions.  Apertures with
1.5\asec\ and 0.75\asec\ radii proved best for the LRIS and PANIC
images, respectively.  The $AB$ magnitudes of each object are recorded
in Table~\ref{tab:Phot}.  These have been corrected for atmospheric
extinction as well as Galactic extinction.  Again, where necessary we
interpolated to find the extinction in the specific bandpasses; the
resulting Galactic extinction values were $\rm A_{G}=0.90$\,mag, $\rm
A_{R}=0.60$\,mag, and $\rm A_{H}=0.13$\,mag.

We attempted to perform photometry on the objects detected in the
NIRC2 LGSAO image, however the complexity of the AO PSF combined with
the close angular spacing of all objects made this task extremely
difficult.  We were unable to recover any trusthworthy values.  Future
improvements in adaptive optics PSF determination would help remedy
photometric measurements of extremely complex sightlines such as this
one.

\begin{deluxetable*}{llllccccc}
\tablecaption{\sc{Galaxy Photometry} \label{tab:Phot}}
\tablewidth{0pt}
\tablehead{\colhead{Object} & \colhead{F555W\tablenotemark{*}} & \colhead{F625W\tablenotemark{*}} & \colhead{F775W\tablenotemark{*}} & \colhead{$r_{\rm Kron}$ (\asec)} & \colhead{$r_s$ ($h^{-1}$ kpc)} & \colhead{LRIS $g$} & \colhead{LRIS $R$} 
& \colhead{PANIC $H$}}
\startdata
$A$  & 23.8 $\pm$ 0.05 & 23.0  $\pm$ 0.05 & 22.6  $\pm$ 0.05 & 1.7 & 4.1 & 23.8 $\pm$ 0.2 & 22.9 $\pm$ 0.2 & 22.4 $\pm$ 0.2 \\ 
$B$  & 26.9 $\pm$ 0.16 & 26.1  $\pm$ 0.05 & 25.2  $\pm$ 0.05 & 0.7 & 2.0 & $>$25.0 & $>$24.0 & $>21.46$ \\
$C$  & 25.4 $\pm$ 0.07 & 24.7  $\pm$ 0.05 & 23.9  $\pm$ 0.05 & 1.2 & 2.8 & 24.8 $\pm$ 0.3 & 24.0 $\pm$ 0.2 & 22.5 $\pm$ 0.2 \\
$G1$ & 24.9 $\pm$ 0.05 & 24.9  $\pm$ 0.05 & 24.7  $\pm$ 0.05 & & & ... & ... & ... \\
$G2$ & 26.2 $\pm$ 0.07 & 26.1  $\pm$ 0.05 & 25.8  $\pm$ 0.05 & & & ... & ... & ... \\
$G3$ & 25.5 $\pm$ 0.06 & 25.2  $\pm$ 0.05 & 24.9  $\pm$ 0.05 & & & ... & ... & $>$25.0 \\
\enddata
\tablecomments{$AB$ magnitudes of every object identified within 5\asec\
  of the afterglow position.  Objects are labeled in
  Figures~\ref{fig:groundPanel} and~\ref{fig:ACS}.  The magnitudes
  have been corrected for Galactic extinction using the values
  reported by \citet{Schlegel:1998}.  Where necessary we interpolate
  to find the extinction in the specified bandpass.  The extinction
  values used were: $\rm A_{555}=0.73$\,mag, $\rm A_{625}=0.63$\,mag, $\rm
  A_{775}=0.47$\,mag, $\rm A_{g}=0.90$\,mag, $\rm A_{R}=0.60$\,mag, $\rm
  A_{H}=0.13$\,mag.}
\tablenotetext{*}{All HST magnitudes for each object are assumed to have a systematic error of at least 0.05 mag which exceeds the statistical uncertainty.}
\end{deluxetable*}

\section{Galaxy Properties Along the Line of Sight Toward GRB 060418}\label{sec:losGalaxies}

The {\it Swift} detection of an optical afterglow associated with
GRB\,060418 triggered several target-of-opportunity spectroscopic
campaigns.  The afterglow was observed soon after the burst with the
Magellan Inamori Kycoera Echelle spectrometer
\citep[MIKE;][]{Bernstein:2003} on the Magellan 6.5\,m Clay telescope
at the Las Campanas Observatory, and with the Ultraviolent and Visual
Echelle Spectrograph (UVES) at the Very Large Telescope (VLT) in
Chile.  In addition to detecting low-ion resonant and fine-structure
transitions at $z=1.490$ (which establish the redshift of the GRB host
galaxy), both groups reported the detection of three strong Mg\,II
absorption systems along the sightline at redshifts of $z=0.603$, 0.656,
and 1.107 \citep{Prochaska:2007b,Ellison:2006}.  The groups measured
similar Mg\,II equivalent widths; we adopt the rest-frame values by
\citet{Prochaska:2007b}: $W(2796)=1.2704 \pm 0.013\,\AA$, $0.9725 \pm
0.010\,\rm\AA$, and $1.8414 \pm 0.023\,\rm\AA$, for the three
absorbers in order of increasing redshift.

While it may seem extraordinary that three strong $W(2796) \gtrsim 1)$
Mg\,II absorbers would exist along one sightline toward a GRB, it has
been shown that many GRB afterglow spectra exhibit a strong Mg\,II
absorber and/or foreground damped \lya\ absorber.  Specifically,
\cite{Prochter:2006} report an apparent factor of four overabundance
of Mg\,II systems along GRB sightlines compared to those along quasar
sightlines.  Furthermore, \cite{Chen:2009} report an overdensity of
galaxies near the afterglow for GRB with spectra showing strong
\ion{Mg}{2} absorption.  These puzzling results remain to be
explained.  Indeed, the on-average higher incidence of absorbers along
GRB sightlines partially motivated this research.  In addition, we
were motivated by the promise of analyzing the stellar properties of
these absorbing galaxies without being hampered by a bright background
quasar, as would be the case in traditional QSO absorption studies.

In the following sections we explain our interpretation of galaxies
$A$, $B$, and $C$, detected in our ground based images, as the three,
strong Mg\,II absorbers known to intervene the sightline at redshifts
of $z=0.656$, 1.107, and 0.603, respectively.  We attribute objects
$G1$, $G2$, $G3$ as part of the host galaxy of the GRB at $z=1.491$,
due to their similar blue colors.

\subsection{Properties of Galaxy $A$}

Figure~\ref{fig:Longslit} shows that the LRIS spectrum of galaxy $A$
displays emission features that are consistent with a star-forming
galaxy at $z=0.6554\pm 0.0002$.  We calculate the total flux in the
[\ion{O}{2}] nebular lines to be $(7.9 \pm 0.3)\times 10^{-18}$
ergs\,cm$^{-2}$\,s$^{-1}$, which corresponds to a star formation rate
of about $\approx 0.2\,\rm M_{\odot}\,yr^{-1}$ using the
\citet{Kennicutt:1998} relation.  We note that the star formation rate
is likely higher than this estimate (but within a factor of two),
because we have not accounted for Galactic extinction or slit losses.
The emission redshift of the galaxy falls at $\delta v = 110 \pm
40$\,\kms\ blueward of the velocity of a strong Mg\,II absorber of
$W(2796)=0.97\pm 0.01$\,\AA\ whose peak optical depth (measured from
\ion{Fe}{2} transitons) lies at $z_{\rm Mg\,II}=0.6560$.  At the
distance of the absorber, the projected angular separation between the
galaxy and the afterglow line of sight, $\Delta\,\theta=3.4$\asec,
corresponds to a physical impact distance of $\rho=16.5\,h^{-1}$\,kpc.
The high spatial resolution image of the galaxy displayed in
Figure~\ref{fig:ACS} exhibits a well-developed stellar disk with an
estimated Kron radius \citep{Kron:1980} of 1.7\asec.  For an
exponential profile, the Kron radius (the first moment of surface
brightness distribution) is twice the disk scale length.  The ACS
image of galaxy $A$ therefore indicates a disk scale length of
$4.1\,h^{-1}$\,kpc at $z=0.6554$, comparable to the scale length of
the Milky Way at the present time.

\begin{figure}
\begin{center}
\includegraphics[scale=0.5]{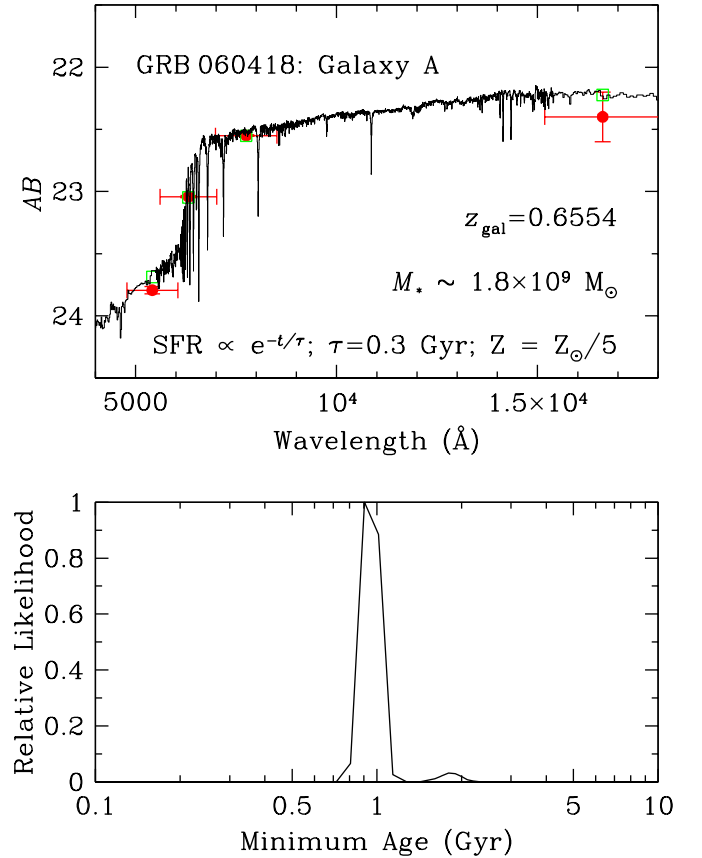}
\caption{Top: Comparison of the observed SED of Galaxy $A$ and the
  best-fit stellar population synthesis model.  The observed
  broad-band photometric measurements are shown in solid points with
  errorbars.  The data were taken through the HST/ACS F555W, F625W,
  and F775W filters and the Magellan/PANIC $H$ filter.  The horizontal
  errorbars indicate the FWHM of each bandpass.  The solid curve
  represents the best-fit synthetic model with the open squares
  indicating the predicted brightness in respective bandpasses.  Based
  on a maximum likelihood analysis, we find that the observed SED is
  best described by an exponentially declining star formation history
  on a 300-Myr characteristic time scale.  Bottom: The likelihood
  function of the minimum age of the underlying stellar population as
  described by the broad-band SED.}
\label{fig:gal_A}
\end{center}
\end{figure}

The precise redshift of the galaxy allows us to examine the underlying
stellar population using the observed optical and near-infrared
colors.  The top panel of Figure~\ref{fig:gal_A} displays the observed
broad-band spectral energy distribution (SED) of galaxy $A$ over the
wavelength range of $\Delta\,\lambda=5000-16,000$\,\AA.  To constrain
the stellar population and star formation history of the galaxy, we
consider a suite of synthetic stellar population models generated
using the \citet{Bruzual:2003} spectral library. We adopt a Salpeter
initial mass function with a range of metallicities from
$0.2\,Z_{\odot}$ to $Z_{\odot}$ and a range of star formation
histories from a single burst model to a tau-model with an
exponentially declining star formation rate (${\rm SFR}\propto
\exp(-t/\tau)$) with $\tau=0.3 - 1$\,Gyr.  We include no dust in our
synthetic spectra.  A maximum likelihood analysis that compares the
data with a grid of models shows that in the absence of dust the
observed SED is best described by an exponentially declining star
formation history on a 300-Myr characteristic time scale, with a
metallicity of $0.2\,Z_{\odot}$.  In addition, the minimum age of the
stellar population is found to be $\sim 1$\,Gyr, as shown in the
likelihood function in the bottom panel.  Given the simple stellar
population synthesis model, we further estimate the total stellar mass
to be $M_*\sim 2\times 10^9\ {\rm M}_\odot$.  We note that dust
extinction and the adopted metallicity are degenerate in synthetic
spectra.  The estimated stellar age and mass are insensitive to the
assumed dust content.

\subsection{Properties of Galaxies $B$ and $C$}

Both galaxies $B$ and $C$ are located closer to the afterglow line of
sight than galaxy $A$.  Both galaxies display relatively redder colors
than what is observed of galaxy $A$, but galaxy $C$ displays a
well-resolved disk structure that is similar in apparent size to
galaxy $A$.  We measure a Kron radius of 1.2\asec\ for galaxy $C$ and
0.7\asec\ for galaxy $B$.  Given our slit orientation and our
wavelength coverage, [O\,II] emission from galaxies $B$ and $C$ should
have been observed if these galaxies are at $z\le 1.4$ modulo very
bright sky lines.  We searched the spectra at the positions of
galaxies $B$ and $C$ for nebular lines and found no line emission.  To
determine the redshifts of galaxies $B$ and $C$, we therefore rely on
their observed colors and the {\it a prior} knowledge of two strong
Mg\,II absorbers with $W(2796)>1.2$\,\AA\ at $z=0.603$ and 1.107 along
the GRB sightline.

\begin{figure}
\begin{center}
\includegraphics[scale=0.5]{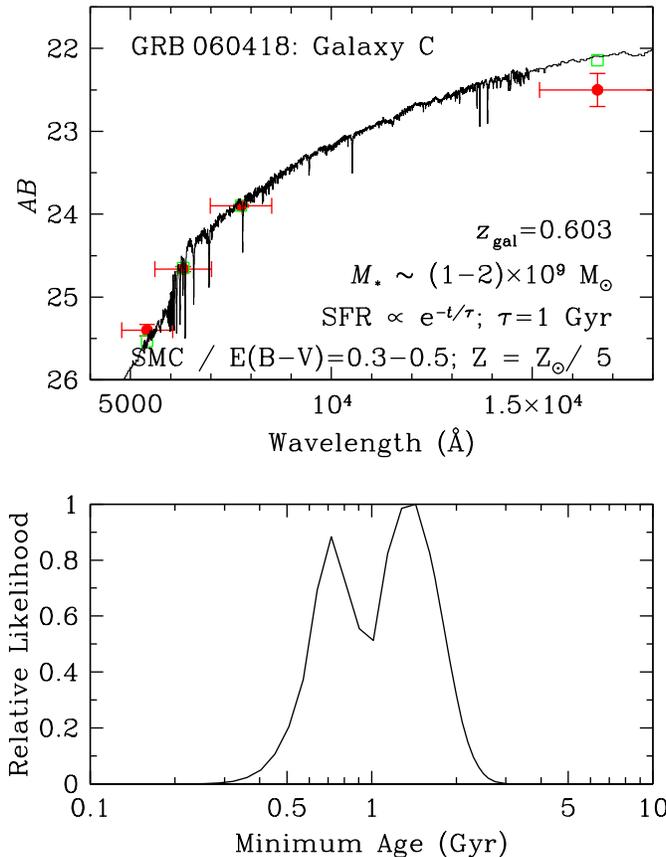}
\caption{Top: Comparison of the observed SED of Galaxy $C$ and the
  best-fit stellar population synthesis model.  The observed
  broad-band photometric measurements are shown in solid points with
  errorbars.  The data were taken through the HST/ACS F555W, F625W,
  and F775W filters and the Magellan/PANIC $H$ filter.  The horizontal
  errorbars indicate the FWHM of each bandpass.  The solid curve
  represents the best-fit synthetic model with the open squares
  indicating the predicted brightness in respective bandpasses.  Based
  on a maximum likelihood analysis, we find that the observed SED is
  best described by an exponentially declining star formation history
  on a 300-Myr characteristic time scale.  Bottom: The likelihood
  function of the minimum age of the underlying stellar population as
  described by the broad-band SED.}
\label{fig:gal_C}
\end{center}
\end{figure}

We argue that galaxy $C$ is responsible for the Mg\,II absorber at
$z=0.603$ for two reasons.  First, the observed optical and
near-infrared colors are consistent with an Sab galaxy at $z\approx
0.6$.  Second, the apparent angular size of galaxy $C$ is comparable
to that of galaxy $A$, at a similar redshift.  At $z=0.603$, the
observed Kron radius indicates a disk scale length of
$2.8\,h^{-1}$\,kpc.  To further understand this galaxy, we examine the
underlying stellar population by considering a suite of stellar
population synthesis models.  Motivated by the observed red color, we
include dust extinction and adopt an SMC extinction law.  We find that
the broad-band SED is best described by an exponentially declining
star formation history with $\tau=1$\,Gyr (Fig.~\ref{fig:gal_C}).  The
minimum age of the stellar population is found to be $\sim 3$\,Gyr, as
shown in the likelihood function in the bottom panel of
Figure~\ref{fig:gal_C}.  The likelihood function also indicates that
the SED can be interpreted nearly equally well by a $\tau=0.3$\,Gyr
model but with a younger age.  In either model, we estimate that a
total extinction of $E(B-V)\approx 0.3-0.5$ is necessary, and the
total stellar mass is $M_*\sim 2\times 10^9\,{\rm M}_\odot$.

Finally, we argue that galaxy $B$ is responsible for the highest
redshift Mg\,II absorber.  Galaxy $B$ has an extremely red color
suggesting a large extinction, which is consistent with the large
depletion found for the strong Mg\,II absorber at $z=1.107$
\citep{Prochaska:2007b}.  A dusty scenario for galaxy $B$ is further
supported by the work of \citet{Ellison:2006}, in which they fit an
absorber at $z=1.118^{+0.004}_{-0.001}$ with a Milky Way extinction
law using $E(B-V)=0.08 \pm 0.01$ and $\rm A_{V}=0.25$.  At $z=1.107$,
the observed Kron radius of galaxy $B$ indicates a disk scale length
of $2\,h^{-1}$\,kpc.  Additional near-infrared spectroscopic
observations of the galaxy may confirm the identification of the
Mg\,II absorber based on the presence of an H$\alpha$ emission line.

The small angular separation beween galaxies $G1$ and $B$,
$\Delta\,\theta\approx 0.6$\asec, suggests that gravitational
magnification may be effective if $G1$ is indeed located behind $B$.
To estimate the effect of lensing, we first estimate the mass of $B$.
Galaxy $B$ is observed to have an unobscured luminosity of $0.4\,L_*$.
Correcting for dust-extinction estimated by \citet{Ellison:2006}, we
derive an intrinsic luminosity of $0.6\,L_*$ for galaxy $B$.  Adopting
the dark matter halo mass-to-light ratio for $z\approx 1$ galaxies
from \citet{zcz+07}, we estimate that the hosting dark matter halo has
a total mass of $M_h\sim 10^{12}\,h^{-1}\,{\rm M}_{\odot}$ with a
virial velocity of $\approx 204$ km s$^{-1}$ at $z=1$.  Adopting a
singular isothermal density profile leads to an Einstein radius of
$\approx 0.4$\asec\ for galaxy $B$, which in turn suggests that $G1$
may have been magnified by $\approx 20$\%.

We summarize the photometric properties of all galaxies within 5\asec\
of the afterglow in Table~\ref{tab:Phot}.  Table~\ref{tab:Summary}
summarizes the properties of these three galaxies having adopted the
galaxy identifications discussed above.  Knowing each galaxy's
redshift as well as its angular separation from the GRB afterglow, we
can now associate the measured gas properties with a physical galactic
radius at which the photo-ionized gas presides.  If we use a simple
model for Mg\,II gas decreasing with galactocentric radius
\citep[e.g.][]{CT:2008}, we would expect galaxy $A$ to exhibit the
smallest Mg\,II equivalent width because its halo was probed at the
largest galactocentric radius of 29\,h$^{-1}$\,kpc.  This
interpretation is consistent with our results.  

We note that additional galaxies are found at angular distances beyond
5\asec, which would have projected distances of $>33$ kpc at $z=0.6$
and $>40$ kpc $z=1.1$ from the afterglow line of sight.  While it is
unlikely that these galaxies would be physically connected to these
strong Mg\,II absorbers (see e.g.\ Bouch\'e et al.\ 2007; Chen \&
Tinker 2008), follow-up spectroscopic survey of these galaxies will
allow us to study in detail the galaxy environment of these strong
Mg\,II absorbers.

\section{Discussion}

The sightline toward GRB 060418, exhibiting six distinct objects all
within 3.5\asec\ of the afterglow's position, highlights the
difficulty of distinguishing morphologically complex host galaxies
from multiple foreground absorbing galaxies at close angular
distances.  The high resolution HST images allow us to resolve
multiple foreground galaxies from the GRB host and attribute objects
$G1$, $G2$, $G3$ as part of the host galaxy of the GRB at $z=1.491$
based on their similar blue colors.  Follow-up campaigns with limited
wavelength coverage, low resolutions, or shallow depths, can easily
lead to mis-identifications, even when $8-10$\,m telescopes are used.
Our Keck/LRIS images taken at $g$ and $R$ band provide an example of
this risk; in these data galaxy $A$ is the only well resolved system,
and its proximity to the reported UVOT position might lead to the
misidentification of the $z=0.656$ absorber as the host.

This complex sightline also reminds us that {\it Swift} XRT positions
should be treated with caution.  In Fig.~\ref{fig:ACS} we show the
revised XRT position and 90\% containment radius in yellow, overlaid
on the HST/ACS image.  This XRT position \citep{Falcone:2006a} just
barely encircles the $G3$ component of the host galaxy and the actual
afterglow position, which is depicted by a pink plus symbol.  If
follow-up observations had proceeded based on the XRT position alone,
any instrument with a field of view smaller than $\sim$8\asec\ across
would have missed the host galaxy altogether, and it is likely that
galaxy $C$ would have been mis-identified as the host.  Thus we
encourage a healthy dose of skepticism when drawing conclusions about
GRB host galaxies in the cases of dark bursts \citep[e.g.][]{cobb08}.
In addition, we warn that bursts without accurate UVOT positions are
especially limited for calculating impact parameters to intervening
galaxies; the angular separation from the XRT position and the actual
afterglow position corresponds to physical distances on the order of
30\,kpc at $z \sim 1$.

As a final note of caution regarding analysis using positions solely
from {\it Swift}, we note that even the UVOT position (depicted by a
green circle in Fig.~\ref{fig:ACS}) can be misleading.  In the case of
GRB\,060418, the UVOT position is nearly coincident with the $G2$
component of the host, while the actual afterglow position is $\sim
0.5$\asec\ to the south.  Since \citet{Bloom:2002} calculated a median
projected offset of $\approx 0.9$ kpc (corrected for $H_{0}=75\,\rm
km\,s^{-1}\,Mpc^{-1}$), corresponding to 0.12\asec\ at $z=1.491$
between afterglows and the flux-weighted centroids of their putative
host galaxies, one may have been tempted to conclude that this
particular GRB occured within the stellar environment of component
$G2$.  However, in this case the positional error of 0.5\asec\
corresponds to $\sim 4.3$\,kpc at $z=1.49$.  This level of
uncertainty, on kpc physical scales, significantly hinders the
interpretation of the local environments of GRBs, and highlights the
need for early-time follow-up images that complement {\it Swift's}
onboard instruments.  For those times when HST is unavailable for
early-time follow-up, it is imperative that we develop a system for
accurate positional determination from the ground.  In this work we
have shown that by using a combination of low-resolution, early-time,
automated observations from 1\,m class telescopes and high-resolution,
deep, late-time observations from $8-10$\,m class telescopes, we can
calculate the afterglow's position relative to the host and foreground
absorbers to within just 0.02\asec, or 170\,pc.\footnote{Our absolute
positional error depends on the positional error of the 2MASS
catalog.}

With such an accurate afterglow position, we can now interpret the
gaseous properties of the host galaxy and the three foreground
absorbers at a galactocentric radius known to better than 200\,pc.
Next, we discuss these gaseous properties and compare them to the
stellar properties.

\begin{deluxetable}{llcccccc}
\tablecaption{{\sc Properties of} MgII {\sc Absorbers} \label{tab:Summary}}
\tablewidth{0pt}
\tablehead{\colhead{Object} & \colhead{$z_{\rm abs}$} & \colhead{$\rho$\tablenotemark{a}} & \colhead{$M_{B}$\tablenotemark{b}} & \colhead{$W(2796)$\tablenotemark{c}} & 
\colhead{$\log[N(\rm FeII)/cm^{-2}]$} \\
\colhead{} & \colhead{} & \colhead{($h^{-1}$ kpc)} & \colhead{$-5$\,log$h$} & \colhead{(\AA)} & \colhead{}}
\startdata
$A$ & 0.656 & 16.5 & $-$18.8 & 0.97  & 13.82 $\pm$ 0.06 \\
$B$ & 1.107 & 8.25 & $-$19.6 & 1.84  & 14.59 $\pm$ 0.08 \\
$C$ & 0.603 & 7.5  & $-$17.2 & 1.27  & $15.67\pm 0.3$ \\
$G$ & 1.490 & 7.16 & $>-$18.4 & 1.93  & 15.22$\pm$0.03  \\
\enddata
\tablenotetext{a}{Separation (in projection) between GRB afterglow and
  MgII absorbers, assuming the listed redshifts.  For $G$, the host, we
  give the separation between the afterglow and the host's
  flux-weighted centroid, calculated from the F775W filter.}
\tablenotetext{b}{Absolute, rest-frame, $B$-band, $AB$ magnitude,
  calculated by interpolating the observed magnitudes between the
  available wavebands.  For the host, the observed $H$-band limiting
  magnitude of component $G3$ translates almost directly to the
  rest-frame $B$-band limit and no interpolation was done.}
\tablenotetext{c}{Rest equivalent width of MgII $\lambda 2796$
  measured in \AA, taken from \citet{Prochaska:2007b}.}
\end{deluxetable}

\subsection{The Host Galaxy of GRB 060418}\label{sec:host}

GRB 060418 occured along a very complex sightline, but by performing
photometric and spectroscopic observations of the objects in this
field we have disentangled three foreground systems from the
complicated host galaxy.  The blue host exhibits a disturbed
morphology, with at least three components seperated by 2\asec.  The
extended system is reminiscent of the $z=1.09$ host of GRB~980613
\citep{dbk00}, the $z=2.04$ host of GRB 000926 \citep{Fynbo:2002}, and
the $z=2.14$ host of GRB 011211 \citep{Jakobsson:2003}.  The system
conforms to our current picture of GRB hosts as blue and irregular
galaxies experiencing a mode of ongoing star formation which is
different from that found in the massive starbursts typified by mid-IR
and sub-mm detections \citep{Le-Floch:2003,Fruchter:2006}.  Most of
the GRB hosts observed to date have $\sim 0.1\,L_{*}$ luminosities
\citep{Chen:2009}.  Using the limit of $H>25.0$\,mag calculated from
our PANIC data, we conclude that this host's rest-frame absolute $B$
magnitude, $M_{B, rest}-5$\,log\,$h$, is greater than $-18.4$\,mag.
Using $M_{B_{*}}-5$\,log\,$h =-20.5$ for blue galaxies at $z_{\rm gal}
\sim 1.3$ \citep{Faber:2007}, this limit on $M_{B, rest}$ corresponds
to a luminosity limit $L \lesssim 0.1\,L_{*}$, consistent with other
GRB hosts.

Nearly all GRB hosts are found to be DLAs, and the GRB-DLA population
has a median value of $\log\,N_{\rm HI} = 21.7$ \citep{Prochaska:2007}.
However, this host resides at $z=1.49$, making the Ly$\alpha$
absorption feature impossible to observe from the ground.  Assuming an
upper limit of $\log\,N_{\rm HI} = 23$, \citet{Prochaska:2007} calculated a
lower limit on the gas phase metallicity of this host, along the
afterglow's sightline, and found $\rm [M/H]>-2.65$.  Unfortunately,
while this metallicity limit is consistent with what we expect for
GRB-DLAs, it does not significantly contribute to our current
understanding of relationships between luminosity, age, dust, or
metallicity.  GRB-DLAs exhibit metallicities with an intrinsic scatter
of $\sim2$\,dex, and none have shown a metallicity $\rm [M/H]<-2.4$
\citep{Prochaska:2007}.

The lack of concrete values for $M_{B, rest}$ and $\rm [M/H]$ thwarts
our aim to sketch a complete picture of this high redshift galaxy,
comparing its stellar and gaseous properties.  Yet in this case, using
our accurate knowledge of the location of the burst with respect to
the multi-component system, we can still significantly increase our
understanding of the burst's local environment and begin to appreciate
what conditions produce GRBs rather than supernova, or other less
energetic star formation products.  This burst was rather unique in
that it did not conform to the usual scenario in which the afterglow
is situated very near to the flux-weighted centroid of its host.
\citet{Bloom:2002} calculated a median projected offset of 0.17\asec\
between afterglows and the flux-weighted centroids of their putative
host galaxies, but the afterglow from GRB\,060418 occured 1.2\asec\
away from its host's flux-weighted centroid.  (The flux-weighted
centroid is just south-west of galaxy $B$.)  Even if component $G1$
has been incorrectly identified as part of the complex host system,
the offset between the afterglow and components $G2$ and $G3$ is still
0.45\asec.  It has been suggested that the tight correlation between
afterglow locations and the locations of the brightest knots in their
host galaxies implies that GRB projenitors are the {\it most} massive
stars - more massive than typical supernova progenitors \citep{wp07}.
Examples such as GRB\,060418 should at least prompt a critical redress
of these claims and consider that a large range of projenitors may
exist.  Similar conclusions have been drawn from examples of GRBs that
may have ocurred in the halos of the host galaxy
\citep{Cenko:2007,Perley:2008} or with large offsets from the most
intense star forming regions within the galaxy \citep{hfs+06}.

Finally, we note that we have entered an exciting time when we can
probe equally sized physical scales by means of high-resolution
imaging and detailed analysis of early-time afterglow spectroscopy.
\citet{Vreeswijk:2007} analyzed UVES spectroscopy taken roughly ten
minutes after this burst.  From the time evolution of the FeII and
NiII excited and metastable populations, they concluded that neutral
gas resides at $1.7 \pm 0.2$\,kpc from the afterglow.  If this neutral
gas was associated with an old stellar population, we could have
resolved it using our ground-based adaptive optics images or HST/ACS.
Note that components $G2$ and $G3$ of the host, which are closest to
the burst's location in projection, are at least 3.5\,kpc away.
Adopting the sensitivity limit of the ACS/F625W images, we derive a
2-$\sigma$ upper limit to the underlying star formation rate (SFR)
intensity of 0.0074 M$_\odot$ yr$^{-1}$ kpc$^{-2}$ over a region of 1
kpc radius at the afterglow's position\footnote{Here we have
temoprarily adopted $H_{0}=70\,\rm km\,s^{-1}\,Mpc^{-1}$ to compare
with the \citet{Vreeswijk:2007} result.}.  For $\log\,N({\rm
H\,I})\approx 21.3$ (the median value of GRB DLAs; see e.g.\
\citealt{jakobsson+06b}), we estimate a surface gas mass density of
$\approx 16$ M$_\odot$ pc$^{-2}$.  The observed SFR intensity limit is
consistent with the expectation of the Schmidt-Kennicutt star
formation law for nearby normal disk galaxies (e.g.\ \citealt{KS98}).
We conclude that the gas observed along this sightline is unrelated to
an old stellar population nor an intense, actively star-forming region
within the galaxy (unless that region is highly obscured).

\subsection{Foreground Absorbing Galaxies}\label{sec:absorbers}

The three galaxies identified through their Mg\,II absorption toward
GRB\,060418, and discovered in emission in this work, offer a unique
opportunity to study the combined stellar and gaseous properties of
high redshift galaxies.  Previous samples of Mg\,II absorption systems
detected toward quasars have long been scrutinized; when searching for
stellar counterparts of absorbing systems near a bright backround
source there is an intrinsic bias toward higher impact parameters and
luminosities.  The fading nature of GRB afterglows enables a
systematic search for the absorbers even at zero impact parameter.

Despite the clear advantages of this exciting alternative to quasar
absorption line studies, forward progress has been slow.  Early-time
spectroscopy is needed to identify the redshifts of Mg\,II absorbers,
late-time deep images are needed to discover the stellar counterparts,
and finally late-time follow-up spectroscopy is needed to confirm the
identifications.  To date, only two Mg\,II absorbers have ever been
spectroscopically confirmed along GRB sightlines.
\citet{Jakobsson:2004} found a $z=0.842$ Mg\,II absorber 1.2\asec\
from the optical afterglow of GRB\,030429, and \citet{Masetti:2003}
discovered a $z=0.472$ Mg\,II absorber $\sim$2\asec\ from the optical
afterglow of GRB\,020405.  Galaxy $A$, in this work, represents the
third such spectroscopically confirmed absorber.

Like the two previously identified, spectroscopically confirmed, MgII
absorbers along GRB sightlines, all three galaxies intervening
GRB\,060418 are considered ``strong'' absorbers, with $W(2796) \gtrsim
1$\,\AA.  It is instructive to compare the properties of galaxies $A$,
$B$, and $C$, to a larger sample of QSO-selected, strong MgII
absorbers.  The three GRB-selected absorbers discussed in this work
should represent an unbiased subset of the larger random sample,
although the possible excess of \ion{Mg}{2} absorbers along GRB
sightlines raises some doubts \citep{Prochter:2006}.  The unobscured
absolute rest-frame magnitudes of each galaxy are summarized in
Table~\ref{tab:Summary}.  If $M_{B_{*}}-5$\,log\,$h =-20.5$, then
these magnitudes correspond to roughly $0.2\,L_{*}$, $0.4\,L_{*}$, and
$0.05\,L_{*}$ galaxies for $A$, $B$, and $C$,
respectively\footnote{Adopting the best fit SMC extinction law from
Figure 5, the extinction corrected absolute magnitude of galaxy $C$ is
comparable to what is found for galaxy $A$.}.  Historically, strong
Mg\,II absorbers at redshifts less than 1 have been found to have
luminosities comparable to $L_*$ \citep{Bergeron:1986,Steidel:1994}.
In a more recent study which involved a statistical analysis of
thousands of stacked images from the SDSS, \citet{Zibetti:2007} found
that strong MgII absorbers ($W(2796)>0.8$ \AA) typically arise in
$\sim 0.5\,L_{*}$ galaxies (see also Kacprzak et al.\ 2008).
Considering that all three galaxies identified in this work have
sub-$L_{*}$ luminosities, it is suggestive that older quasar
absorption studies were biased by the quasar's blinding light, and
included mis-identifications.

This is in part confirmed by a recent study of \citet{CT:2008}, who
have shown that the equivalent widths of Mg\,II absorbers at a known
galactocentric radius can be described by an isothermal density
profile of Mg$^{+}$ ions.  Remarkably, the equivalent widths and
impact parameters of galaxies $A$, $B$, and $C$ fit very well with
this latest result, which was compiled using \ion{Mg}{2} systems first
detected along quasar sightlines \citep[See Fig.\,5 of ][]{CT:2008}.
In addition, galaxies $A$, $B$, and $C$ extend this relationship to
significantly smaller galactocentric radii -- a parameter space that
is precluded by the quasar in traditional absorption studies.

Finally, we note that the large ion abundances observed in the Mg\,II
absorbers suggest that these are likely damped \lya\ absorption (DLA)
systems of $\log\,N({\rm H\,I})\ge 20.3$ \citep[e.g.][]{wgp05}.  In
particular, the Mg\,II absorber at $z=0.603$ contains $\log\,N({\rm
Fe\,II})=15.67$ \citep{Prochaska:2007b}.  A solar abundance of the
absorbing gas without dust depletion would imply an underlying total
neutral gas column density of $\log\,N({\rm H\,I})=20.2$, and a 0.1
solar metallicity would imply $\log\,N({\rm H\,I})=21.2$.  Similarly,
the Mg\,II absorber at $z=1.107$ is expected to contain at least
$\log\,N({\rm H\,I})=20.4$ given the observed Zn abundance
\citep{Prochaska:2007b}.  Our imaging and spectroscopic survey of the
sightline toward GRB\,060418 has therefore unveiled two new DLA
galaxies at $z\apll 1$.  It demonstrate the the unique sensitivity in
probing neutral gas cross section selected DLA galaxies using GRB
afterglows.  Known properties of galaxies $B$ and $C$ not only confirm
that the luminosity distribution of DLA galaxies is consistent with a
neutral gas cross-section selected field galaxy population
\citep[e.g.][]{Chen:2003}, but also provide unobstructive views of new
DLA galaxies.

\acknowledgements 

We thank A.\ Gal-Yam for pointing out the possibility of galaxy $G1$
being lensed by $B$, and an anonymous referee for constructive
comments on the paper.  H.-W.C. acknowledges partial support from
HST-GO-10817.01-A and an NSF grant AST-0607510.  L.K.P. was supported
by the NSF Graduate Research Fellowship and by NASA/Swift grants
NNG06GJ07G and NNX07AE94G.  J. X. P. is partially supported by
NASA/Swift grants NNG06GJ07G and NNX07AE94G and an NSF CAREER grant
(AST-0548180).  Some of the data presented herein were obtained at the
W. M. Keck Observatory, which is operated as a scientific partnership
among the California Institute of Technology, the University of
California, and the National Aeronautics and Space Administration
(NASA).  Some of the data presented herein were obtained in part at
the Magellan telescopes, a collaboration between the Observatories of
the Carnegie Institution of Washington, University of Arizona, Harvard
University, University of Michigan, and Massachusetts Institute of
Technology.  PAIRITEL is operated by the Smithsonian Astrophysical
Observatory (SAO) and was made possible by a grant from the Harvard
University Milton Fund, a camera loan from the University of Virginia,
and continued support of the SAO and UC Berkeley. The PAIRITEL project
are further supported by NASA/Swift Guest Investigator grant
NNG06GH50G. We thank Dan Starr, Cullen Blake, Mike Skrutskie, Emilio
Falco, Andrew Szentgyorgyi, Ted Groner, and Wayne Peters for making
PAIRITEL possible.

\bibliographystyle{../apj}
\bibliography{../lindsey_refs}

\end{document}